# GENDER DATA 4 GIRLS?: A POSTCOLONIAL FEMINIST PARTICIPATORY STUDY IN BANGLADESH

Isobel Talks, University of Oxford, isobel.talks@education.ox.ac.uk

**Abstract:** Premised on the logic that more, high-quality information on majority world women's lives will improve the effectiveness of interventions addressing gender inequality, mainstream development institutions have invested heavily in gender data initiatives of late. However, critical empirical and theoretical investigations into gender data for development policy and practice are lacking. Postcolonial feminist theory has long provided a critical lens through which to analyse international development projects that target women in the majority world. However, postcolonial feminism remains underutilised for critically investigating data for development projects. This paper addresses these gaps through presenting the findings from a participatory action research project with young women involved in a gender data for development project in Bangladesh. Echoing postcolonial feminist concerns with development, the 'DataGirls'[1] had some concerns that data was being extracted from their communities, representing the priorities of external NGOs to a greater extent than their own. However, through collaborating to develop and deliver community events on child marriage with the 'DataGirls', this research demonstrates that participatory approaches can address some postcolonial feminist criticisms of (data for) development, by ensuring that gender data is enacted by and for majority world women rather than Western development institutions.

## 1. INTRODUCTION

Over the last ten years, mainstream development actors such as the UN, Gates Foundation and the World Bank, have all launched gender data for development initiatives. The argument made by these proponents of the 'gender data revolution' (Fuentes and Cookson, 2020) is, in the words of Melinda Gates (2016), that "we cannot close the gender gap without first closing the data gap". This treatise is based on the belief that a "paucity of data hampers effective policy-making" in the first instance, and "hinders effective monitoring and accountability" in the second (Chattopadhyay, 2016: p.4). In other words, having the right data on women and girls' lives should mean that development organisations can design projects that address gender inequality more effectively and evaluate these initiatives more accurately.

Yet despite the fact that numerous "powerful international development institutions have called for a 'gender data revolution'" and the implications that this has for women and girls' lives, as of yet there has been little critical academic study of this phenomenon, with Fuentes and Cookson's (2019: p.881) paper the notable exception. In particular, empirical studies of development projects claiming specifically to collect digital data in order to close the 'gender data gap' and the 'gender gap' are lacking. Critical data for development (D4D) studies is an emerging field (Taylor and Broeders, 2015), as is data feminism (D'Ignazio and Klein, 2020). The former draws on critical development theory and concepts to analyse data for development projects, and the latter draws on critical feminist theoretical concepts to deepen our understanding of how data projects perpetuate intersectional gendered inequalities. Yet there is little overlap between the two nascent fields, with data for development projects specifically seeking to address gendered power relations receiving negligible critical attention. Postcolonial feminist theorists have been providing critical insight into the intersectional power relations inherent in development projects that target women in the majority world for decades. However, as of yet no academic research has been carried out taking a postcolonial feminist theoretical approach to the 'gender data revolution' in international development.

---

[1] Pseudonym used for anonymity purposes

This paper addresses these gaps by presenting findings from a participatory action research study that collaborated with young women engaged in a gender data for development project in Bangladesh. The young women involved in the 'DataGirls' project felt that the data they were collecting should be shared with the community rather than extracted for use by external NGOs. However, whilst exploring ways to make the 'DataGirls' findings accessible to the community, it also emerged that the young women wanted to share different information than that which they had been tasked to gather by the 'DataGirls' project. These findings echo postcolonial feminist critique of mainstream international development interventions as perpetuating misrepresentations of majority world women and being aligned to the interests of development actors, thus furthering neocolonial and patriarchal power relations (Kothari, 2002; Syed and Ali, 2011). Through working in participation with some of the 'DataGirls' to share information on child marriage with their communities, the gendered subject they felt was most vital to address, the research detailed in this paper demonstrates that participatory approaches can enable gender data for development projects to be enacted by and for majority world women, thus responding to postcolonial feminist concerns, albeit within structural constraints.

## 2.     POSTCOLONIAL FEMINISM AND DEVELOPMENT

Postcolonial feminists have addressed weaknesses in both postcolonial and feminist theory. Postcolonial theory has often neglected the ways in which "gender and sexuality are necessarily imbricated in colonialism and its legacies" (Chambers and Watkins, 2012: p.297). In response to this oversight, some postcolonial feminists have sought to bring women's experiences of colonialism to the fore (Chaudhuri and Strobel, 1992, Jayawardena, 1992), whilst others have deconstructed the ways in which notions of femininity and masculinity in the colonial era shaped, and were shaped by, imperialism (McClintock, 1995; Sinha, 1995; Stoler, 2002).

As well as extending postcolonial studies, postcolonial feminist theorists have also highlighted the limitations of Western liberal feminism. For example, Spivak (1985: p.243) demonstrated how through upholding the notion of a transhistorical, singular 'womanhood' Western "feminist criticism reproduces the axioms of imperialism". When white second-wave feminists proclaimed 'sisterhood is global' they negated the ways in which racism intersects with gendered oppression (Combahee River Collective, 1978; Lorde, 1979; Moraga and Anzaldua, 1983; Smith, 1983). Furthermore, Mohanty (1984: p.353) argued that where white Western feminists did try to represent women in the majority world, they often did so by collapsing their diverse experiences into a sole oppressed and vulnerable "third world woman", a trope that through juxtaposition props up the equally untrue stereotype of Western women as "secular, liberated and having control over their own lives".

These postcolonial feminist ideas have been very influential in critical development studies. Kothari (2002: p.48) notes that development studies is "replete with texts that present as homogenous diverse groups of people and practices, such as books on … gender relations in Sub-Saharan Africa". Syed and Ali (2011) argue that present-day gender and development research and projects are the continuation of the colonial trope of the 'white woman's burden', in which white Western feminists play the role of expert and 'saviour' to the majority world female 'victim', whilst ignoring indigenous women's subject position and sense of agency. In Abu-Lughod's (2002: p.789) words,

"projects of saving other women depend on and reinforce a sense of superiority by Westerners, a form of arrogance that deserves to be challenged". After all, as Trinh (1989) notes, the power that women in the West have to study and intervene in the majority world is not often possible the other way around. Sardar (1999) agrees that it is in the ability to define, represent and theorise about the majority world that neo-colonial power truly lies.

In response to these postcolonial feminist critiques, some academics working in the majority world on international development related issues have sought to find other ways of knowing and being beyond Eurocentric narratives, for example through making space for majority world women to speak and represent themselves, so that the "material complexity, reality and agency of Third World women's bodies and lives" can be re-established (Mohanty, 2003: p.510). For some, such as Miles and Crush (1993), this has meant engaging in life history research with women in the majority world "as a way of recovering hidden histories, contesting academic androcentrism, and reinstating the marginalised and disposed as makers of their own past". They have also tried to develop new ways to 'speak with' rather than 'to' or 'for' the people with whom one is engaged in research", leading to the rise in popularity of participatory development approaches (McEwan, 2001: p. 102). For example, Segebart adopted a participatory approach in her research into the monitoring of municipal development plans in the Brazilian Amazon as a way to "respond to feminist (and) postcolonial claims of decolonising and breaking down hierarchies of knowledge production" (Schurr and Segebart, 2012: p. 148).

It is this last point, that engaging in participatory development work can respond to postcolonial feminist criticisms of international development, that this paper builds upon. As Asiedu (2012: p.1200) argues, postcolonial feminist theory and critique is largely absent from ICT4D scholarship, despite the fact that it can "offer important ideas to address gender and ICT issues facing the global south". More recently, Narayanaswamy (2016) has utilised postcolonial theory to critically examine the Knowledge for Development (K4D) paradigm, including the way that the discourse constructs women in the majority world as possessing a 'knowledge deficit' that requires training and access to ICTs led by Northern-based development experts. Dé et al. (2018) also draw from the work of postcolonial feminist Gayatri Spivak's ideas about the subaltern voice to advocate for using local languages rather than defaulting to English as the main operating language in ICT4D. However, there remains a wide scope for new work on ICTs and development that draws upon the wider postcolonial feminist canon. For example, just as there is little scholarship on 'gender data for development' projects in general, as "feminist and postcolonial scholarship and data studies remain as largely unconnected fields with little cross-fertilisation" (Leurs (2017: p.134), there is also an absence of work analysing the rise of gender data for development from a postcolonial feminist perspective.

Therefore, what follows is a postcolonial feminist analysis of a participatory action research intervention into a gender data for development project with young women in Bangladesh. This paper provides a detailed account of the participatory process that was undertaken, the findings of this process and also reflections upon the extent to which taking a participatory approach to gender data for development addressed postcolonial feminist concerns about development.

## 3. METHODOLOGY

The empirical data drawn upon for this paper was collected and analysed as part of the author's PhD. The methods utilised included six months of participant observation of an NGO project in Bangladesh, as well as 77 interviews, two participatory workshops and three community events with some of the stakeholders engaged in the project. This research was carried out from April to June, and September to December 2019. The interview and participatory workshop data were translated and transcribed with the help of two research assistants. These transcripts, as well as the participant observation research diary, were analysed in line with Carney's (1990 in Miles and Huberman, 1994: p.2) 'ladder of analytical abstraction', moving from summarising and packaging the data through In Vivo and evaluative coding, through to repackaging and aggregating the data by collating and synthesising emerging themes, before constructing an explanatory framework through mapping these themes against the literature to draw conclusions.

 'DataGirls' - the project on which this research was carried out - was an international government funded intervention designed by a UK head-quartered multinational NGO and delivered in partnership with an implementing Bangladeshi partner NGO. Commencing in 2018, and lasting for two years, this project entailed recruiting and training 48 young women (aged 16 – 19) from four different regions of Bangladesh to become smartphone-based digital data collectors. These young women were given smartphones with a research application uploaded onto them, through which they were set various data collection tasks. These data collection tasks were carried out for five other multinational NGOs engaged in work with girls in Bangladesh, and the topics were decided by these organisations in discussion with the 'DataGirls' NGO staff based in the UK and Bangladesh. The 'DataGirls' were paid for their data collection work, and were awarded with a Market Research Society qualification in digital research.

## 4.  FINDINGS

This paper presents the findings of a participatory action research collaboration with 10 of the 'DataGirls' in Bangladesh. We collaborated to address one of the concerns that the young women had raised regarding 'DataGirls': the fact that the gender data they gathered on behalf of external NGOs was not shared with the community from whom it was extracted. Through working together to try to address this and 'close the loop' of the gender data, it also emerged that the young women felt there were other topics that they would prefer to share information on with their communities rather than those that they had been assigned to collect. Engaging in this participatory process with the 'DataGirls' therefore brought two key critical questions to the fore: 1) Who is gender data for?, and 2) Who is gender data by?. If gender data for development projects are not by and for majority world women, but are instead by and for international NGOs, then this perpetuates the colonial and patriarchal power relationships that postcolonial feminists have criticised international development for furthering. As such, this paper provides empirical evidence that strengthens the argument that participatory approaches can go some way towards responding to postcolonial feminist concerns about (data for) development. However, as the discussion section will argue, participatory development practices can go some way towards responding to postcolonial feminist critique but cannot alone dismantle the structural global power relations which continue to oppress majority world women and create the conditions for data-led 'development' in the first place.

### 4.1 Gender data for whom?

A concern that many of the 'DataGirls' raised was the fact that the data that they gathered on behalf of the NGOs was not fed back to them or the wider community:

> Me: "Do you know about the results of the [project] research?"
>
> Lamia[2]: "No, all the information goes straight to the office"
>
> Me: "Do you want to know?"
>
> Lamia: "It would be good, because I would know how well I had done through them … how much all of the things are improving"

As this interview excerpt with Lamia, one of the 'DataGirls', demonstrates, the young women involved in this gender data for development project wanted to know the results of the research that they had carried out. In part they wanted to see the data so that they could 'know how well' they had done and therefore assess their own performance as a digital data collector. However, they also wanted to know the findings so that they could know 'how much all of the things are improving' – i.e. so that they too could understand what topics people in their community felt positively about, as well as issues that needed attention. As one of the other 'DataGirls', Rakiba told me:

---

[2] Pseudonyms have been used for all participant names for anonymity purposes

> "Sister, we cannot get the right idea about the problem without research. Suppose there is shortage of drinking water in the community but after research we can know that the main problem is food scarcity - so research is best"

Whilst Lamia, Rakiba and many of the other 'DataGirls' were interested in having the findings of the data that they had gathered returned to them, they did not only want this data to be shared with themselves alone. The 'DataGirls' also felt that it was important for the wider community to have access to the information gathered during the data collection, so they too could have their 'awareness raised'. The 'DataGirls' told me that one of the main positive outcomes of the 'DataGirl' project for the wider community was that those that they collected data from had gained useful information whilst being interviewed:

> "We talk to them regarding the things we were trained on. We explain the topic to them. For instance, there were two topics called Food and Nutrition 1 and Food and Nutrition 2. So, girls in our locality didn't know much about food and nutrition. When we were made to research on it, we spoke to the girls about these things, and so they got to know these things from us. And then we found out information about them that we hadn't known. Added together, the whole of it was great. When we sent the recordings to our seniors, they understood the whole picture" (Fahima)

However, as the 'DataGirls' only gathered data from a few people from their local community, selected by the project staff in line with their sampling criteria, they told me that they also wanted the 'DataGirls' research topics to be shared more widely with those living in their local area:

> "The main purpose of TEGA is to raise awareness and help the girls. So, in this case if the parents and elderly people are given training on this context then I think it would be helpful" (Anika)

> "If a meeting was held with the parents in the village and topics like child marriage, eve teasing or other such topics were discussed, they would also realise that girls just don't sit at home anymore they are able to go forward. They can study and then get jobs. Through these kind of meetings awareness will grow amongst the parents and they'll encourage their daughters to study" (Razia)

This call from the 'DataGirls' to have the data that they had gathered on behalf of NGOs about girls and gender equality issues fed back to them and their communities, rather than remaining solely in the hands of these external organisations for their organisational uses, is in line with debates taking place within the data for development literature (e.g. Thatcher et al. 2016; Albornoz et al. 2019; Mann, 2017). As Narayanaswamy (2016) asks in her postcolonial feminist critique of extractive 'knowledge for development' practices, if knowledge is power then "why should only the Northern technical expert hold knowledge?".

Building on this, I decided to organise a participatory action workshop with the young women in one of the four 'DataGirls' networks to explore further, and hopefully develop some solutions to, this issue of 'closing the gender data' loop. My research assistant and I travelled to Northern Bangladesh in October 2019 to meet with 10 of the 'DataGirls'. Originally there had been 12 in this network, but one (Tasnina) had given birth recently, and another had married and left the project earlier in the year. The location for the workshop was a local government hall equidistant from the three villages where the 10 'DataGirls' lived. The reason that this network was selected, instead of the other three networks, was due to the fact that this concern about the need to share the findings of the data collection with themselves and the wider community had been raised most often by the young women in this network. The workshop lasted from 9.30 am to 4 pm.

In line with the aim of participatory research, which is that the participants rather than the researchers should take the lead (Manzo and Brightbill, 2007), the workshop consisted of various different participatory activities designed so that the young women could take centre stage whilst myself and my research assistant played the role of "conduits, channelling perspectives and voices which would otherwise remain silent" (Gillies and Alldred, 2012: p.49). One of these activities involved making posters on the 'DataGirls' research topics. The 'DataGirls' in this network had collected data on three different topics (girls' access to economic opportunity, knowledge of digital financial services, and pregnant adolescents' nutritional knowledge) for three different NGOs. Prior to the workshop my research assistant had translated the final reports, written by the NGO for the NGO clients whom the data was gathered for, from English to Bangla. A copy of these was given to each of the 'DataGirls' for them to read and keep. In three groups the 'DataGirls' created colourful posters exploring these findings, before presenting them to the rest of the group. In the group discussion following the presentations the young women said that the findings in the final reports matched well with what they had remembered from their interviews with the research participants. They also felt strongly that these results should be shared with the rest of the community.

This further confirmation that the 'DataGirls' wanted to share the findings of the data they had gathered back with their communities from whom it was gathered then progressed to the next stage of the participatory action workshop – the 'DataGirls' deciding what data they would like to share.

**4.2 Gender data by whom?**
The next part of the participatory workshop focused on this question – what information did the 'DataGirls' most want to share with their communities? The young women all wrote on a post-it note the one topic that they would most like to share with their community. They stuck their post-it notes to a big piece of paper and this was held up so that everyone could see. 'Child marriage' was voted for the most times, followed by 'Girls Education'. One young woman put 'Child labour' and one put 'Dowry'. What stood out from this exercise was that the topics which the 'DataGirls' had chosen were not those that they had been collecting data on for the external NGOs, but were other topics that they felt were more important to share:

> Tasnia: We all want [the project] to improve even more … if some more work could be done on child marriages …
>
> Me: What sort of work?
>
> Tasnia: I mean also awareness like parents and everyone becoming a little more aware - I think it would be improved.
>
> Me: What sort of work on awareness?
>
> Tasnia: Awareness that girls should not be married off at a young age, that they should study, help their parents, they should work and gain independence. So that they can lead their lives independently … explaining it to them and, like, maybe conducting a training for the parents or through a meeting, explaining it to them.

The fact that the young women had chosen different subjects, such as child marriage, rather than the ones that the NGO had tasked them to collect data on reflects postcolonial feminist critiques of the misrepresentation of women in the majority world. Postcolonial feminist scholar Chandra Talpade Mohanty (1988:334) in her seminal piece 'Under Western Eyes' rails against the way that Western feminist texts have "discursively colonize[d] the material and historical heterogeneities of the lives of women in the third world, thereby producing/re-presenting a composite, singular "Third World Woman". As geographer Joni Seager (2016) points out:

> "women in poor countries seem to be asked about 6 times a day what contraception they use …. But they are not asked about whether they have access to abortion. They are not asked about what sports they like to play"

Rather than be (mis)represented by more powerful others, however, the 'DataGirls' were able to choose collaboratively the information about girls that they wanted shared with their communities during this participatory process. Through a group discussion the 'DataGirls' decided that they would all like to work on one topic together, and that they would select the most voted for topic – child marriage. They also chose community events as the method to disseminate information by, to which the local people and also the chairpersons of each area would be invited. Crucially, the young women wanted these events to take place three times, once in each village so that each community could benefit equally. The workshop ended with us agreeing that we would all meet again to decide on and prepare the content of these community events.

This second participatory workshop took place in November 2019. The day began with participatory diagramming, with the 'DataGirls' writing down what they wanted their communities to know about child marriage, including their own personal views and experiences, or those of people that they knew. Two of the 'DataGirls' present had been married before the age of 18, and shared their own experiences with the rest of the group:

> "I married into a good family so I didn't face too many difficulties. I think it's good that I got married because I was experiencing a lot of eve teasing, but that stopped after I got married. And because I married into an educated family I was able to stay in school, but that's not possible for most girls when they get married" (Moriam)

> "I am still studying too so I am happy … but I have more responsibilities now, like looking after my in-laws" (Farida)

Tansina, the 'DataGirl' who had recently had a baby and so had not been able to attend, had also married before she was 18. As the discussion continued, the rest of the group said that this was a more typical case of child marriage – she had been having an 'affair' with her now husband, so her father had insisted that she married him. However, after the marriage her husband and in-laws began treating her badly, including restricting her freedom of movement, and they did not want her to participate in 'DataGirls' anymore, or continue her education. The group concluded that early marriage is different for different young women, but that these issues were common in most cases they had seen in their communities. They agreed that they wanted to share some of these experiences in their community events, highlighting the challenges young women in their villages faced. We worked together to create a powerpoint on the project laptop that the District Officer of the 'DataGirls' programme had brought to the session. This would be projected onto a screen for the community engagement. The 'DataGirls' selected key information to share, and included a 4 minute World Bank video from YouTube called 'Our Daughters' which tells the story of a young woman in Bangladesh being forced to marry and drop-out of school against her will. In their village groups the 'DataGirls' also chose one of them to present a speech written together during the session. They also asked if the 'DataGirls' District Officer, me, and whichever local officials who attended would also give short speeches.

The 'DataGirls' concluded that the community events turned out to be a success. They were happy with the turnout (55, 65 and 25 attendees respectively) and that the attendees were a mixture of ages and gender identities. In two of the villages this was the first community event there had ever been on child marriage. In the village where there had been workshops on child marriage before, the 'DataGirls' said that they had not been well received in comparison to their own sessions. When I asked why they felt this to be the case they said that this was because the previous events had been one-way lectures, that had no room for open discussions. By comparison, their events had had time for questions and debate following their presentations and speeches. Other aspects that the 'DataGirls' said had made their events a success was the use of digital technology, which they felt engaged people to a greater extent, the inclusion of male as well as female community members, the fact that they had shared good quality information in a personal and engaging way and that they had given participants a chance to share their views and opinions, rather than just being spoken to. Only one of the three village chairmen attended, but the 'DataGirls' said that there were positives and negatives to this. On the positive side, the chairman's attendance meant that he had heard local people's views, and had made a speech against child marriage which the 'DataGirls' felt would have a big influence on child marriage rates. On the other hand, the two events without the presence of the local chairman had livelier, more extensive community debates, which the 'DataGirls' felt were very important, and meant the events had a "good hype" and would be talked about for days to come.

## 5.     DISCUSSION AND CONCLUSIONS

This paper has provided empirical evidence that suggests that taking a participatory approach to gender data for development can address some longstanding postcolonial feminist critiques of international development. Participatory research emerged out of the desire to "break the monopoly on who holds knowledge and for whom social research should be undertaken" (Fine, 2008: p.215), aiming to do research 'with' rather than 'on' people, producing findings that not only directly benefit the participants but also fairly represent them (Cahill, 2007). Resonating with postcolonial feminist concerns about international development interventions being led by and for development actors, thus perpetuating misrepresentations of majority world women as well as neo-colonial and patriarchal power relations, the 'DataGirls' also felt that the project did not gather the gender data they felt was most important to them and their lives, and that it extracted information from their communities rather than sharing knowledge equitably. Engaging in the participatory action research process together not only enabled these issues to be discussed and brought to the fore – it also meant that the 'DataGirls' could work together to gather information on child marriage, the subject they felt was most critical to them and their lives, and to develop and deliver community events to share this 'gender data' with those in their local area. As such, based on these findings, it can be concluded that as a participatory approach shifts gender data for development so that it becomes led by and for majority world women, rather than by and for Western development actors, it can address some postcolonial feminist critiques of (data for) development.

Yet it would be naïve to claim that a short-term, small scope participatory action project such as this provides the simple, instrumental resolution to neocolonial and patriarchal power relations. As Cahill (2007: p. 299) writes "we need to be wary of broad applications of the term 'participation' because it often masks tokenism and the illusion of consultation that may, in fact, advance dominant interests". Whilst I facilitated this participatory research with the 'DataGirls' with the best of postcolonial feminist intentions, in practice it cannot be denied that it replicated some of the problematic aspects of mainstream international development that we sought to disrupt. For example, the participatory process not only provided space for the 'DataGirls' to enact their agency to address some of their concerns with the programme – it also provided findings for my PhD, thus arguably furthering the long history of Westerners profiting from the testimonies of majority world women (Trinh, 1989). As Heeks (2017: p.5) has argued data justice is unavoidably structural. It is impossible for postcolonial feminist data for development, no matter how participatory, to exist in a neocolonial, patriarchal world. However, despite these limitations this paper has provided an empirically substantiated critique of mainstream gender data for development practices, and demonstrated the extent, however limited, to which participatory gender data that is by and for majority world women can address these concerns. These findings therefore contribute to the ongoing debates regarding the role of data and development in bringing about a gender equitable and just world, with implications for future development research and practice.

## REFERENCES AND CITATIONS